\documentstyle[preprint,aps,fleqn]{revtex} 

\newfont{\bi}{cmbxti10 scaled\magstep2}
\begin{document}
\draft

\title{Quantum transport in a multiwalled carbon nanotube.}
\author{L. Langer, V. Bayot, E. Grivei, J.-P. Issi.}
\address{Unit\'e de Physico-Chimie et de Physique des
Mat\'eriaux,
Universit\'e Catholique de Louvain,\\
Place Croix du Sud 1, B-1348 Louvain-la-Neuve, Belgium}
\author{J.P. Heremans, C.H. Olk.}
\address{Physics Department, General Motors Research, Warren, MI 48090, USA}
\author{L. Stockman \cite{stock}, C. Van Haesendonck, Y. Bruynseraede.}
\address{Laboratorium voor Vaste-Stoffysika en Magnetisme,
Katholieke Universiteit Leuven, Celestijnenlaan 200 D, B-3001 Leuven,
Belgium}
\date{\today}
\maketitle

\begin{abstract}
We report on electrical resistance measurements of an individual
carbon nanotube down to a temperature $T$=20~mK.
The conductance exhibits a $\ln~T$ dependence and saturates at
low temperature. A magnetic field applied perpendicular to
the tube axis, increases the conductance and produces aperiodic
fluctuations. The data find a global and coherent interpretation in
terms of two-dimensional weak localization and universal
conductance fluctuations in mesoscopic conductors.
The dimensionality of the electronic system is discussed in
terms of the peculiar structure of carbon nanotubes.
\end{abstract}


With the discovery by Iijima\cite{1} of carbon nanotube (CN) structures,
a new class
of materials with a reduced dimensionality has been introduced. The
nanotubes
are made of coaxial graphitic cylinders. Each cylinder can be visualized as
the conformal mapping of a two-dimensional (2D) honeycomb lattice onto its
surface. Theoretical calculations predict that structural parameters, i.e.
diameter, helicity and number of concentric cylinders, strongly
influence the band structure and hence the electronic properties of CNs
\cite{4}.
This implies that a CN can either be a metal, a semimetal, or an insulator,
depending on structural parameters. It has also been predicted
that the presence of a magnetic field will strongly affect the band
structure near the Fermi level \cite{7}.

Up to now no electrical resistance measurements on individual nanotubes have
been reported. Our group succeded to measure a micro-bundle
(total diameter around 50 nm) of multiwalled CNs \cite{9}. At higher
temperature ($T>$ 1K) we observed a typical semimetallic behavior,
consistent
with the simple two-band model for semimetallic graphite. Song
{\it et al.} reported on a large CN bundle exhibiting a behavior similar to
that of disordered semimetallic graphite \cite{8}.

In this letter we report on the first electrical resistance measurements
performed on an individual CN. In zero magnetic field we observe a
logarithmic decrease of the conductance with decreasing temperature
at high temperature, followed by a
saturation
below $T \approx 0.3$~K. In the presence of a magnetic
field,
applied perpendicular to the tube axis, a pronounced and positive
magnetoconductance (MC), i.e. an increase of conductance with increasing
field, is observed. The temperature dependence of the conductance in a
magnetic field can be described consistently within the framework of the
theory for 2D weak localization (WL) \cite{10}. The effect of the
magnetic field on the density of states
predicted by Ajiki and Ando \cite{7} is also observed. The mesoscopic
dimensions
of the CN result in the appearance of reproducible fluctuations of the MC
with respect to magnetic field.
These fluctuations show an amplitude
and a temperature dependence consistent with the presence of universal
conductance fluctuations (UCF) \cite{11}. Our conductance measurements
strongly support the idea that isolated multiwalled nanotubes
behave as disordered mesoscopic 2D graphite sheets.

The carbon nanotubes were synthesized using the standard carbon
arc-discharge
technique \cite{12}. High resolution TEM images of our CN materials
reveal a few nanotube bundles and many single tubes with an
average diameter around 18 nm.

In order to contact a nanotube, the growth material is dispersed on an
oxidized Si wafer covered with an array of large
square gold pads. After evaporation of a thin
gold film, a few layers of a negative electron resist, $\omega$-tricosenoic
acid, is
deposited using the Langmuir-Blodgett (LB) technique. The scanning
tunneling microscope (STM) is then used
to locally expose the resist from the CN towards the predefined
gold pads \cite{14,15}. The unexposed parts of the LB film are dissolved in
ethanol and the unprotected parts of the gold film are removed by Ar ion
milling. The result of the STM lithography is an electrically connected CN.
High resolution TEM investigations of CNs
with comparable diameter were used to
check that the CNs are not damaged by the ion milling process.

AC electrical
conductance measurements at 15 Hz were performed in the mixing chamber
of a dilution refrigerator in a magnetic field, $B$, perpendicular to the
tube axis. We have been able to attach electrical contacts to four similar
multiwalled
CNs. All samples reveal the presence of a pronounced positive MC on which
reproducible fluctuations are superimposed. Here, we focus on the results
for one particular CN, having a length $L \approx 800$ nm between the two
voltage probes and a diameter $d \approx 20$ nm, as determined by atomic
force microscopy. Three contacts were attached to our CN sample, which
allowed us to check directly that the contact resistances were negligible
compared to the total longitudinal resistance. The observed impedance
was in the ohmic regime for currents ranging from $10^{-11}$ A to
$10^{-9}$ A, and did not present any capacitive component.

In Fig.~\ref{fig:3} we present the conductance $G$ as a function of
temperature for the CN. The conductance clearly shows a $\ln T$ dependence
followed by a saturation at low temperature, while the MC is positive at
all temperatures. The saturation of $G$ occurs at
higher temperatures in the presence of a magnetic field.
A $\ln T$ dependence can either be indicative of weak localization \cite{17}
or disorder enhanced electron-electron interactions \cite{18} in 2D systems,
or can be the signature of a Kondo anomaly due to the presence of magnetic
impurities \cite{19}.
The presence of a Kondo effect is very unlikely, since this effect has
never been observed in graphitic materials even containing
detectable amounts of magnetic impurities \cite{20}. Moreover,
spectrographic analysis revealed that our CN material contains less than
10 ppm (detection limit) of magnetic impurities \cite{21}. The data in
Fig.~\ref{fig:3} also rule out the presence of predominant disorder
enhanced electron-electron interactions since they should exhibit an
opposite behavior, i.e. a positive MC and a saturation of the
low-temperature conductance which is independent of the magnetic field
\cite{10}.

We will now show that our results are consistent with the presence of 2D WL
effects in the concentric graphite cylinders forming the CN. The WL effects
result from the enhanced backscattering probability for the electrons due
to the constructive interference between the partial electron waves
travelling back to their original position along time-reversed paths
\cite{10}.
Since the constructive interference is destroyed by inelastic scattering
(by other electrons or phonons), the backscattering becomes much
more pronounced at lower temperatures, where the inelastic scattering
time $\tau_{in}(T) \propto T^{-p}$, with $p$ depending on the dominant
inelastic scattering mechanism.

At the lowest temperatures, the spin scattering at magnetic impurities,
with a characteristic scattering time $\tau_{s}$, will also contribute to
the destruction of the interference effects \cite{10}.
The spatial extent of the
interference effects is limited to the phase coherence length
$L_{\phi}=\sqrt{D\tau_{\phi}}$, where $D$ is the
elastic diffusion constant and the phase coherence time $\tau_{\phi}$
describes the combined effect of the inelastic and magnetic scattering:
$\tau_{\phi}^{-1}(T) = \tau_{in}^{-1}(T)+2\tau_{s}^{-1}$.
For 2D materials the WL results in a reduction of the
conductance which depends logarithmically on the relevant length scale
$L_{\phi}(T)$. This implies that the WL causes a logarithmic decrease of
$G$ in the highest temperature range, where $\tau_{in}(T) \ll \tau_{s}$,
and a
saturation of $G$ at the lowest temperatures, where
$\tau_{in}(T) \gg \tau_{s}$. This is exactly the behavior which is observed
in Fig.~\ref{fig:3} at $B=0$  \cite{41}.

Assuming that the spin-orbit scattering is weak \cite{20}, the WL theory
can also account for the pronounced positive MC in the presence of a
perpendicular magnetic field. Indeed, the magnetic field also contributes
to the dephasing of the interference effects when the Landau orbit
size $L_B=\sqrt{\hbar /eB}$ becomes smaller than the phase coherence
length $L_{\phi}(T)$.
For a sufficiently high magnetic field, i.e. when
$L_B\ll L_{s}=\sqrt{D\tau_{s}}$, the applied magnetic field governs the
saturation of the conductance at lower temperatures.
This is in agreement with the observation in Fig.~\ref{fig:3} that
the saturation of $G$ occurs at higher temperatures when increasing the
magnetic field.

In contrast to crystalline graphite, the random stacking of graphene sheets
in turbostratic graphite induces a 2D behavior of the electron system
with respect to the WL effects \cite{20}.
The cylindrical structure of CNs produces a similar situation since the
relative positions of carbon atoms in adjacent cylinders of a CN are
randomized due to their specific
geometry. Consequently, each individual graphite cylinder behaves as an
individual intrinsic 2D system.
In the presence of 2D WL effects, the conductance of the CN, contributed by
$n$ cylinders in parallel, is related to the 2D conductivity $\sigma$ via
$G=\sigma \pi nd/L$ and is given by \cite{10}:
\begin{equation}
G(T)=G_o + \frac{e^2}{2 \pi^2 \hbar} \frac{n\pi d}{L} \ln
\left[1+\left( \frac{T}{T_c(B,\tau_s)}\right)^p \right].
\label{eq:2}
\end{equation}

The quantitative analysis of the
conductance using Eq.~\ref{eq:2}, provides us with
values $p$=1, $n$=4 and $T_c$ at different $B$ (Fig.~\ref{fig:3}).
The value $p$=1 is in
agreement with previous experiments \cite{20}
and indicates that the dominating inelastic scattering mechanism is
likely to be disorder enhanced electron-electron scattering in 2D systems
\cite{32}.
The value
for $n$ is likely to
be smaller than the total number of concentric cylinders which are present
in the
multiwalled CN. Indeed, we make electrical
contacts mainly to the outer cylinder and
the coupling with the inner cylinders is very weak due to the very large
anisotropy
of the graphitic material. Moreover, non-conducting cylinders may also be
present.

The observation of weak localization requires that elastic scattering
dominates inelastic scattering. This generally limits the observation of
weak localization as the temperature increases. However, contrary to
common metals, weak localization was observed up to relatively high
temperatures in 2D turbostratic graphite \cite{20}.
This comes from the basic differences between common
metals and graphitic materials, including CNs, where important parameters
such as screening, Debye temperature and Fermi energy are drastically
different.

The WL theory predicts that when $T\gg T_{c}$ the conductance becomes
independent of $B$ \cite{10}.
The data in Fig.~\ref{fig:3} show that this is not the case and that there
is an additional contribution to the MC
of the CN which is independent of temperature.
Ajiki and Ando \cite{7} have predicted the formation of Landau states when
a magnetic field is applied perpendicular to the nanotube axis.
In particular, a Landau level
should form at the crossing of the valence and conduction bands,
thereby increasing the density of states at the Fermi level, and hence
increasing the conductance.
The resulting positive MC is expected to be temperature independent
as long as $k_{B}T$ remains smaller than the width of the Landau level.
Previous results obtained on a
CN bundle \cite{9} revealed a behavior consistent with these predictions.
In the present work, we attribute the additional temperature independent
positive MC to the
same mechanism. Both 2D WL and ``Landau level'' (LL) contributions to the
MC can be
separated as illustrated in Fig.~\ref{fig:3} for $B$ = 7 T.

The temperature $T_c$, which is obtained from Eq.~\ref{eq:2}
corresponds to $L_{s}\approx L_{in}(T_c)$ for low magnetic fields and to
$L_{B}\approx L_{in}(T_c)$ for sufficiently high fields.
The data for $B$ = 14 T allow us to determine that
$L_{in}\approx L_{B}\approx 7 $ nm at $T=T_{c}\approx$ 1.5 K.
Since $L_{in}=\sqrt{D\tau_{in}}\propto T^{-p/2}$, the zero-field
data give $L_{s}\approx L_{in}\approx 20$ nm at $T=T_{c}\approx$ 0.3 K.
Thus, at the lowest temperatures $L_{\phi}$ is smaller than but comparable
to $L$, implying
that we may observe interference phenomena related to the mesoscopic
dimensions of the CN.

In Fig.~\ref{fig:4} we show the details of the MC behavior at different
temperatures. At low temperature, reproducible,
aperiodic fluctuations of the conductance appear, which are superimposed
on the positive background caused by the WL and Landau level effects.
Similar fluctuations have often been observed \cite{11} in disordered metals
and semiconductor structures and have been related to the
tuning of sample-specific electron interference processes by the
perpendicular magnetic field (Aharonov-Bohm effect) \cite{25}.
These fluctuations are well known as ``universal conductance fluctuations''.
The rms amplitude of the fluctuations $\delta G$ reaches the
``universal'' value $\mbox {rms}[\delta G]\approx e^{2}/h$ as soon as
the sample size $L$ becomes smaller than both the phase-coherence length
$L_{\phi}$
and the thermal diffusion
length $L_{T}~=~\sqrt{\hbar D/k_{B}T}$ \cite{31}.
When either $L_{\phi}$ or $L_T$ becomes smaller than $L$, the shorter
of these two length scales governs the amplitude of the fluctuations.

We will now show that the results presented in Fig.~\ref{fig:4} are
consistent with the
stochastic self-averaging of the UCF which occurs in samples with
$L\gg L_{\phi}$. Fig.~\ref{fig:5} shows the temperature dependence of
$\mbox {rms}[\delta
G]$ which reaches a constant amplitude at low temperature and
decreases following a $T^{-1/2}$ power law at higher temperature.
In order to calculate $\mbox {rms}[\delta G]$, a smooth background $G_B$
(dashed line in Fig.~\ref{fig:4}) has been subtracted from the MC data so
that $\left<\delta G\right> = \left< G-G_B\right>=0$ (the brackets refer to
an average over the magnetic field).
The saturation in the temperature dependence of the fluctuations occurs
at a temperature
$T_c^*\approx$ 0.3 K (Fig.~\ref{fig:5}) which, in contrast to the
$T_c$ values discussed above,
is independent of $B$. Since $T_c^*$ coincides with the temperature
$T_c(B=0)$ at which $L_{\phi}$ saturates, we conclude that $L_{\phi}$ is
the relevant length scale for both 2D WL and UCF.

Since for our CN sample $L_{\phi}<\pi d<L$, it can be divided into
$L/L_{\phi}$ conductors in series and $n\pi d/L_{\phi}$ conductors
in parallel, each of them producing fluctuations of the order $e^2/h$.
The statistical self-averaging of the UCF results in fluctuations with
rms amplitude \cite{33}:

\begin{equation}
\mbox {rms}[\delta G]=0.61 \frac{e^2}{h} \left( \frac{n\pi d}{L_{\phi}}
\right)^{1/2} \left( \frac {L_{\phi}}{L}
\right)^{3/2}.
\label{eq:3}
\end{equation}

With the geometrical parameters for our CN, Eq.~\ref{eq:3} predicts
$\mbox {rms}[\delta G]\approx 8.5\: 10^{-3} e^2/h$ below $T_c^*$, in good
agreement with the observed saturation value
$\mbox {rms}[\delta G]\approx 9\: 10^{-3} e^2/h$.
Eq.~\ref{eq:3} also predicts that above $T_c^*$
$\mbox {rms}[\delta G]\propto L_{\phi}\propto T^{-1/2}$
which is indeed observed up to $T$ = 10 K.

We conclude that the electrical transport in a multiwalled carbon
nanotube is governed by typical electron interference effects occuring in
disordered conductors with a reduced dimensionality. The fact that
nanotubes behave as disordered conductors is consistent with the study of
annealing effects on the conduction electron spin resonance \cite{34}.
Despite the very small diameter of the nanotube, the cylindrical structure
of the honeycomb lattice gives rise to a two-dimensional electrical
conduction process.
The temperature and magnetic field dependences of both the weak
localization effects and the universal conductance fluctuations can be
explained consistently in terms of the existing theoretical models,
provided the typical cylindrical geometry of the carbon nanotube
is taken into account.

The authors are much indebted to P.A. Lee, J.-C. Charlier, X. Gonze,
L. Filipozzi, J.-F. Despres and L. Piraux for fruitful discussions
and suggestions.
The work in Belgium has been supported by the Concerted Actions and
Inter-University Attraction Poles programs.
V.B and C.V.H. acknowledge the financial support of the Belgian National
Fund for Scientific Research.


\begin{figure}
\caption{Electrical conductance as a function of temperature at the
indicated
magnetic fields. The solid lines are fits to Eq.~\protect \ref{eq:2}
with parameters $p$=1, $n$=4 and $T_c$=0.3, 1.1 and 1.5 K at
$B$=0, 7 and 14 T respectively. The dashed line separates the contributions
to the magnetoconductance of the Landau levels (LL) and the weak
localization (WL).}
\label{fig:3}
\end{figure}

\begin{figure}
\caption{Magnetic field dependence of the magnetoconductance at different
temperatures.}
\label{fig:4}
\end{figure}

\begin{figure}
\caption{Temperature dependence of the amplitude of $\delta G$
for three
selected peaks (see arrows in Fig.~\protect \ref{fig:4}) as well as $\mbox
{rms}[\delta G]$.}
\label{fig:5}
\end{figure}


\begin{thebibliography}{10}
\bibitem{41}At very low temperatures, extremely small amounts ($<$ppm) of
magnetic impurities can give $\tau_s \ll \tau_{in}$  \cite{10}.
\end{thebibliography}
\end{document}